\documentclass[11pt]{article}
\usepackage[]{graphicx}
\usepackage{amssymb}
\usepackage{epstopdf}
\usepackage{hangcaption}
\title{Mission design of LiteBIRD}
\author{$^a$T. Matsumura, $^b$Y. Akiba, $^c$J. Borrill, $^a$Y. Chinone, $^d$M. Dobbs, \\
$^e$H. Fuke, $^f$A. Ghribi, $^a$M. Hasegawa, $^a$K. Hattori, $^g$M. Hattori, $^a$M. Hazumi, \\
$^f$W. Holzapfel, $^b$Y. Inoue, $^g$K. Ishidoshiro, $^h$H. Ishino, $^b$H. Ishitsuka, \\
$^i$K. Karatsu, $^j$N. Katayama, $^e$I. Kawano, $^h$A. Kibayashi, $^h$Y. Kibe, \\
$^a$K. Kimura, $^a$N. Kimura, $^k$K. Koga, $^p$M. Kozu, $^l$E. Komatsu, $^f$A. Lee, \\
$^e$H. Matsuhara, $^k$S. Mima, $^e$K. Mitsuda, $^m$K. Mizukami, $^a$H. Morii, \\
$^g$T. Morishima, $^m$S. Murayama, $^n$M. Nagai, $^a$R. Nagata, $^m$S. Nakamura, \\
$^o$M. Naruse, $^m$K. Natsume, $^e$T. Nishibori, $^j$H. Nishino, $^e$A. Noda, $^i$T. Noguchi, \\
$^p$H. Ogawa, $^a$S. Oguri, $^q$I. Ohta, $^k$C. Otani, $^f$P. Richards, $^e$S. Sakai, \\
$^a$N. Sato, $^e$Y. Sato, $^i$Y. Sekimoto, $^b$A. Shimizu, $^e$K. Shinozaki, $^e$H. Sugita, \\
$^a$T. Suzuki, $^f$A. Suzuki, $^a$O. Tajima, $^n$S. Takada, $^r$S. Takakura, $^e$Y. Takei, \\
$^a$T. Tomaru, $^i$Y. Uzawa, $^e$T. Wada, $^b$H. Watanabe, $^e$N. Yamasaki, \\
$^a$M. Yoshida, $^e$T. Yoshida, $^e$K. Yotsumoto}
\date{\footnotesize{$^a$High Energy Accelerator Research Organization (KEK), Tsukuba, Ibaraki, Japan \\
$^b$The Graduate University for Advanced Studies (SOKENDAI), Hayama, Kanagawa, Japan \\
$^c$Lawrence Berkeley National Laboratory, Berkeley, CA 94720 USA \\
$^d$McGill University, Montreal, Quebec, Canada \\
$^e$Japan Aerospace Exploration Agency (JAXA), Sagamihara, Kanagawa, Japan \\
$^f$University of California, Berkeley, Physics, Berkeley, CA 94720 USA \\
$^g$Tohoku University, Sendai, Miyagi, Japan \\
$^h$Okayama University, Okayama, Okayama, Japan \\
$^i$National Astronomical Observatory of Japan (NAOJ), Mitaka, Tokyo, Japan \\
$^j$Kavli Institute for The Physics and Mathematics of The Universe (IPMU), \\ The University of Tokyo, Kashiwa, Chiba, Japan \\
$^k$Terahertz-wave Research Group, RIKEN, Sendai, Miyagi, Japan \\
$^l$Max-Planck-Institut fur Astrophysik, Karl-Schwarzschild Str. 1, 85741 Garching, Germany \\
$^m$Yokohama National University, Yokohama, Kanagawa, Japan \\
$^n$University of Tsukuba, Tsukuba, Ibaraki, Japan \\
$^o$Saitama University, Saitama, Japan \\
$^p$Osaka Prefecture University, Osaka, Japan \\
$^q$Kinki University, Higashiosaka, Osaka, Japan \\
$^r$University of Osaka, Osaka, Japan \\ }
$email: tomotake.matsumura@kek.jp$ }
\begin{document}
\maketitle

%\vspace{-0.5cm}
\abstract{
LiteBIRD is a next-generation satellite mission to measure the polarization of the cosmic microwave background (CMB) radiation. On large angular scales the B-mode polarization of the CMB carries the imprint of primordial gravitational waves, and its precise measurement would provide a powerful probe of the epoch of inflation. The goal of LiteBIRD is to achieve a measurement of the characterizing tensor to scalar ratio $r$ to an uncertainty of $\delta r=0.001$. In order to achieve this goal we will employ a kilo-pixel superconducting detector array on a cryogenically cooled sub-Kelvin focal plane with an optical system at a temperature of 4~K. We are currently considering two detector array options; transition edge sensor (TES) bolometers and microwave kinetic inductance detectors (MKID). In this paper we give an overview of LiteBIRD and describe a TES-based polarimeter designed to achieve the target sensitivity of 2~$\mu$K$\cdot$arcmin over the frequency range 50 to 320~GHz.
}

\section{Introduction}
Measurements of the cosmic microwave background (CMB) play a crucial role in modern cosmology. In particular, precise measurement of the polarization of the CMB is the key to probing beyond the last scattering surface, and a community-wide effort is now underway to detect the primordial B-mode signal and measure the critical ratio of the CMB's tensor to scalar fluctuations, $r$. Current and planned ground-base and balloon-borne experiments employ $1000\sim10000$ detectors to provide the sensitivity needed to detect $r\sim0.01$. While these will constrain some inflation models, a next-generation CMB satellite is needed to probe this signal to $r\sim0.001$ in order to test the dominant large-single-field/slow-roll inflation models experimentally. A satellite provides the ideal environment for CMB polarization measurements due to the absence of atmospheric contamination and the ability to achieve full sky coverage, and a number of studies of such next-generation CMB satellites are now in progress~\cite{epic_im,core,litebird,pixie,prism}.

LiteBIRD is a small satellite project which is currently considering two detector technologies -- transition edge sensor (TES) bolometers and microwave kinetic inductance detectors (MKID). In this paper we present an overview of LiteBIRD and describe a focal plane design that employs TES bolometers, which are both widely used in current suborbital CMB experiments and have recently been selected for the SAFARI instrument on SPICA~\cite{safari_spica}.

\section{Mission overview}
\subsection{Science goal and design concept}
The science goal of LiteBIRD is to measure the primordial B-mode polarization of the CMB with the precision necessary to constrain the tensor to scalar ratio to $\delta r<0.001$. The mission design is driven by this goal; from Fisher matrix analysis we have derived our required sensitivity to be $3~\mu$K$\cdot$arcmin, and have then set a target sensitivity of $2~\mu$K$\cdot$arcmin~\cite{litebird}.  With the assumptions of 70~\% fractional sky coverage and 2~years of observations at 100~\% efficiency, the corresponding required and target noise equivalent temperature ($NET$) over the entire array of bolometers in the focal plane are $NET_{arr} = 1.7~\mu K\sqrt{s}$ and $1.1~\mu K\sqrt{s}$ respectively.

\subsection{Frequency coverage}
CMB polarization measurements require broadband frequency coverage in order to separate contaminating foreground emissions (e.g. dust and synchrotron) from the CMB signals. Katayama and Komatsu (2011) show that foreground subtraction using a template method can reduce the foreground residuals from synchrotron and dust down to a bias $\delta r < 0.0006$ given three observation bands at 60,100 and 240~GHz with sensitivities of 0, 2, and 0~$\mu K$arcmin~\cite{katayama_2011}.
% check - sensitivity of zero?!

\begin{figure}
\begin{center}
   \includegraphics[width=5cm,angle=0]{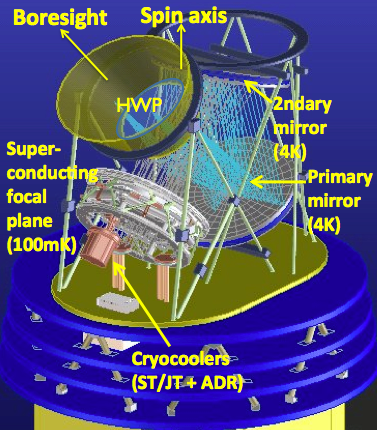} 
   \includegraphics[width=10cm,angle=0]{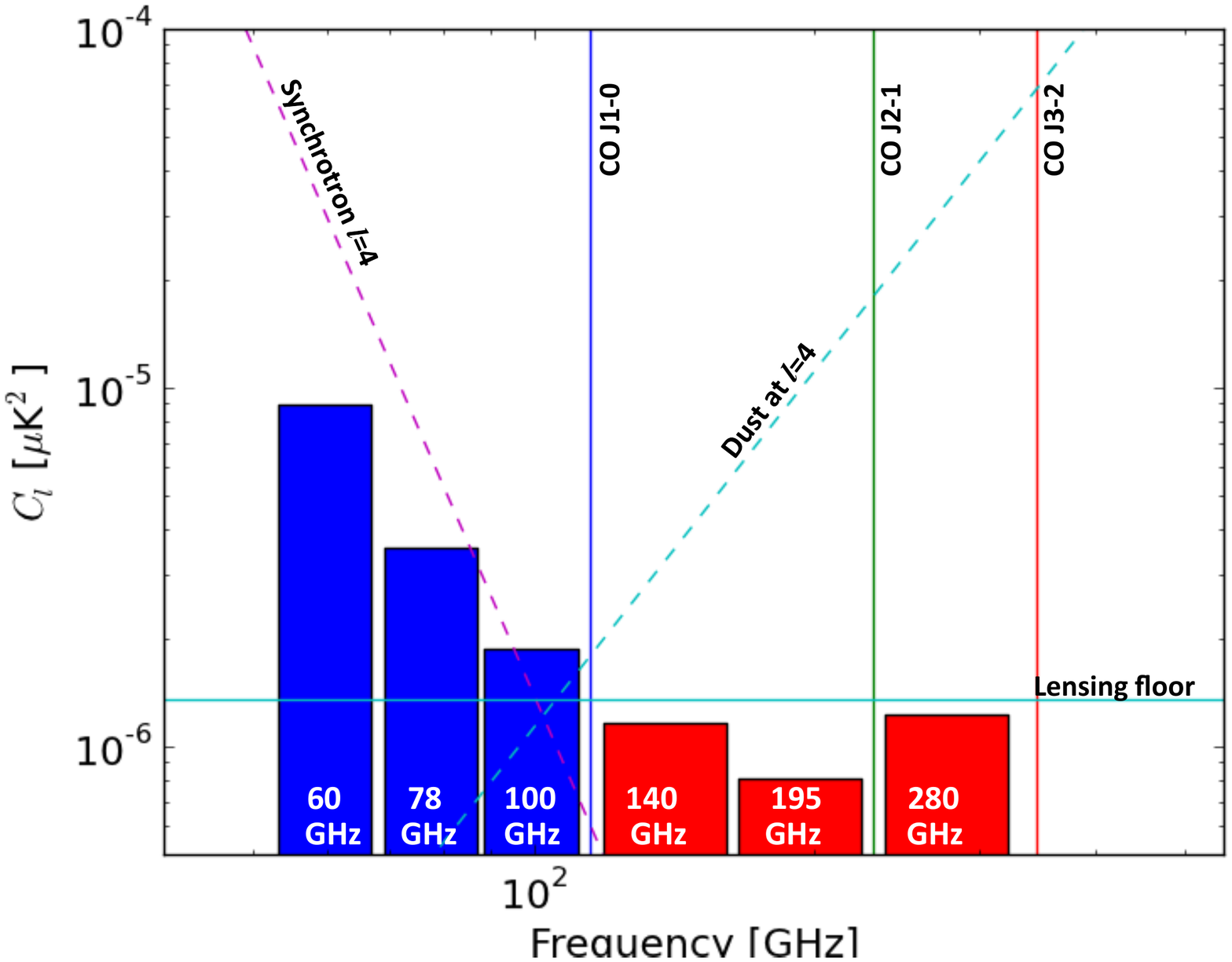}
\hangcaption{ {\footnotesize Left panel: A schematic overview of LiteBIRD. Right panel: The LiteBIRD band locations and the sensitivities in each band; the lowest and highest three bands are each measured by a single lens-coupled pixel simultaneously. The synchrotron and dust emissions are from Dunkley et al. evaluated at $l=4$ with the assumption of a 2~\% residual in the map domain after foreground cleaning~\cite{dunkley_2009}.}\label{fig:bandcoverage}}
\end{center}
\end{figure}

These studies provide a guidance for the choice of observational frequencies as they should span the range 50 to 270 GHz. Assuming that each observing frequency has a bandwidth of $\Delta\nu/\nu \sim30\%$, and that we also avoid the CO lines, then a natural choice of band centers is 60, 78, 100, 140, 195 and 280~GHz as shown in Fig.~\ref{fig:bandcoverage}. The use of six, rather than just three, bands allows us to check internal consistency. In this scheme we generally consider that the 100 and 143~GHz bands are the CMB channels, and the lower and higher two bands are for synchrotron and dust monitor channels respectively. We are also considering the option of distributing the band centers in each pixel in order to increase the effective number of bands. We also have an option to locate a band lower than 60~GHz if it becomes necessary as we learn more about polarized foregrounds. 

\subsection{Detector}
LiteBIRD currently has two detector options, TES bolometer and MKID; given their widespread use, in this paper we focus on the TES bolometer focal plane. We choose the option of using a lens-coupled multichroic TES bolometer, with single lens being coupled to the two polarization states at multiple frequencies simultaneously. In our study, we assume that a single lens measures two polarizations in three colors, and thus a single lens pixel corresponds to six bolometers. This type of bolometer (albeit with only two colors per lens) is being employed by POLARBEAR-2~\cite{suzuki_2012,polarbear2}.  For the MKID option, Karatsu et al. addresses the use of MKID for LiteBIRD~\cite{karatsu}.

For a satellite platform it is essential to make optimal use of the limited instrument size and mass. While it might therefore seem beneficial to increase the number of bands in a single pixel even further, we limit ourselves to three bands per pixel for two reasons. One is that the lens size determines the feed beam and it is ideal to match this to $F/\#$ of the telescope. When a single pixel is used for single-mode multi-color beam coupling, the optimal pixel size depends on the observing wavelength. Thus, the benefit of using a multichroic pixel is reduced if its total frequency coverage becomes too broad. Note though that if we chose to increase the number of bands by increasing their density we still have the option of using more than 3 bands per pixel without increasing the overall bandwidth for a single pixel. The other consideration is that the typical lens material is silicon or alumina, and the lens surface requires an anti-reflection (AR) coating. A three-band AR coating only requires a bandwidth of $\Delta \nu/\nu\sim1$, as has been demonstrated in a lab with moderate effort~\cite{suzuki_2012}. Therefore we opt to split the six bands into two categories, and populate the focal plane with low frequency pixels (LFP) at 60, 78 and 100~GHz and high frequency pixels (HFP) at 140, 195 and 280~GHz.

The nominal focal plane temperature (i.e. the bath temperature of a bolometer) is assumed to be 100~mK. In Section~\ref{sec:discussion} we discuss the sensitivity penalties when the bath temperature is instead 300 or 560~mK.

\subsection{Optical system}
In order to estimate the detector sensitivity it is important to consider the total loading of a bolometer from all the optical components properly. In this paper, we assume that LiteBIRD uses a crossed Dragone telescope with an aperture stop, primary and secondary mirrors, and focal plane. The aperture diameter is set to be 300~mm and an achromatic half-wave plate is placed at this aperture to modulate the incident polarization angle. The effective focal length is 1100~mm, and the corresponding $F/\#$ is 3.6. The aperture, achromatic half-wave plate, and mirrors are assumed to be isothermal. Table~\ref{tab:componentlist} shows all the components that we have taken into account. In Section~\ref{sec:discussion} we study the impact of the temperature of the aperture and mirrors on the sensitivity. 

\begin{table}[tb]
\begin{center}
  \begin{tabular}{c|c|c|c}
    Source & Temperature [K] & Emissivity & Efficiency \\ 
    \hline     \hline
    CMB & 2.725 & 1 & 1\\
    \hline
    Achromatic half-wave plate & 4, (2, 7, 10, 30) & 0.1 & 0.99 (AR)\\
    \hline
    Aperture & 4 (2, 7, 10, 30) & 1 & $1-\epsilon_{s}$\\
    \hline
    Primary and secondary mirrors& 4, (2, 7, 10, 30) & 0.005 & 1 \\
    \hline
    Infrared filter & 1 & 0.1 & 0.95 \\
    \hline
    Lens & 0.1 (0.3, 0.56) & 0 & 0.99 (AR)\\
    \hline
    Antenna and micro-strip related & 0.1 (0.3, 0.56) & N.A. & 0.73 \\
  \end{tabular}
\hangcaption{{\footnotesize The optical components are listed with their assumed temperature, emissivity and efficiency. The numbers in parenthesis are the optional temperatures considered for the focal plane and the mirrors and baffle. The parameter $\epsilon_s$ is the spillover efficiency at the aperture. We assume 7 layers of AR coating on the 7 layer sapphire achromatic half-wave plate, though the hardware implementation is yet to be demonstrated.  % check - is this correct? 
}\label{tab:componentlist}}
\end{center}
\end{table}

\section{Focal plane design}
The detector noise is estimated following Mather and Richards~\cite{mather_1982,richards_1994}. The NET is computed by assuming three contributions -- photon noise, phonon noise and readout noise. We further evaluate the noise level as a function of the pixel size, i.e. the edge taper at the aperture for both LFPs and HFPs. 

The photon noise is computed from the optical loading based on the list of sources in Table~\ref{tab:componentlist}. The phonon noise is computed from the same sources and the thermal conductance of the TES leg taken from Mather~\cite{mather_1982}. We take $n=1$ for the 100~mK bath temperature and $n=3$ for bath temperatures above 100~mK, where $n$ is in the thermal conductivity as $\kappa\sim T^{n}$. We choose a factor of 3 as the ratio of the TES saturation power to the nominal optical loading, where the nominal loading is taken from the brightest terrestrial source (excluding the Sun, Earth and Moon). The 2.725~K CMB monopole is the dominant sky signal for all except the 280~GHz channel, where it is Jupiter due to the higher flux and smaller beam. The total noise from the readout chain is assumed to be 7~pA/$\sqrt{Hz}$ as the current equivalent noise~\cite{hannes_thesis}. % check - I'm not sure I understand this sentence!

\begin{figure}
\begin{center}
   \includegraphics[width=14.cm,angle=0]{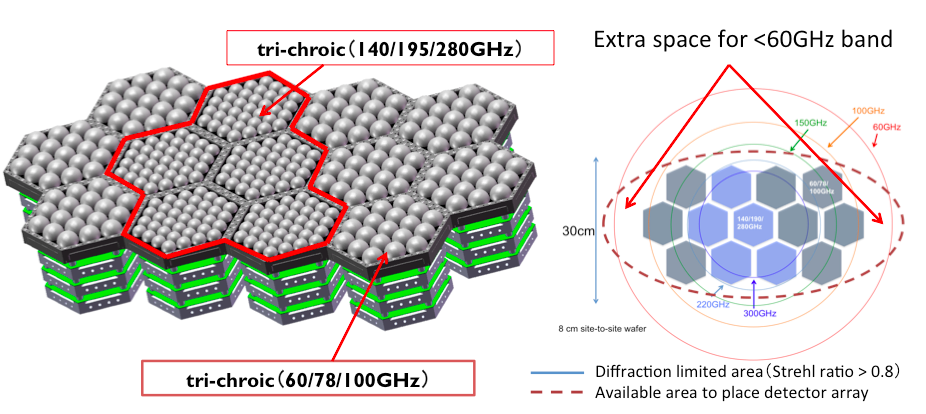}
\hangcaption{{\footnotesize Left: The focal plane layout for LiteBIRD. Right: the diffraction limited area of the crossed Dragone telescope.}\label{fig:lbfp}}
\end{center}
\end{figure}

Figure~\ref{fig:lbfp} shows the focal plane layout. The five central wafers are populated with HFPs and the eight outer wafers with LFPs. The corresponding sensitivity is summarized in Table~\ref{tab:sensitivity}. We employ an 8~cm wide hexagonal wafer in order to locate the detector array within the diffraction limited area of the focal plane. The pixel diameter is chosen in such that the aperture edge taper is about -10~dB for the 100 and 140~GHz bands. The total number of detectors is 2022 and the signal is read by a SQUID and digital frequency domain multiplexing readout system with a multiplexing factor of 64. The power required for the readout is taken to be 2~W per SQUID and thus the total required power for the readout system is less than 100~W~\cite{matt_dfmux1,matt_dfmux2}.

The Planck satellite has reported a high rate of energy deposition on bolometers at L2 due to cosmic rays~\cite{planck_cosmicray}. While addressing this issue is beyond the scope of this paper, we note that this effect has to be carefully studied for the use of wafer-based TES bolometers in space. A high cosmic ray flux can be a significant source of glitches and can correspondingly decrease the observing efficiency and/or complicate post data analyses. When the energy deposition is too high for too long, the TES bolometer has to be re-biased and significant observing time can be lost.

\begin{table}[tb]
\begin{center}
  \begin{tabular}{c|c|c|c|c|c|c|c}
     {\footnotesize Band} & {\footnotesize Beam$^\dagger$} & {\footnotesize NET} & {\footnotesize Pixel \#}  & {\footnotesize Wafer \#} & {\footnotesize Bolometer \#} & {\footnotesize NETarr} & {\footnotesize Sensitivity} \\ 
     {\footnotesize GHz}  & {\footnotesize [arcmin]} & {\footnotesize [$\mu K\sqrt{s}$]} & {\footnotesize per wafer}  &  &  & {\footnotesize [$\mu K\sqrt{s}$]} & {\footnotesize [$\mu K$arcmin]} \\ 
    \hline     \hline
    60 &  75& 99 & 19 & 8 & 304 & 5.7 & 10.3 \\
    78 &  58 & 62 & 19 & 8 & 304 & 3.6 & 6.5 \\
    100 & 45 & 45 & 19 & 8 & 304 & 2.6 & 4.7 \\
    \hline
    140 & 32  & 40 & 37 & 5 & 370 & 2.1 & 3.7 \\
    195 & 24 & 33 & 37& 5 & 370 & 1.7 & 3.1\\
    280 &16 & 40 & 37 & 5 & 370 & 2.1 & 3.8\\
    \hline
    Total &  & & 168 & 13 & 2022 & 1.0 (1.6$^{\dagger\dagger}$) & 1.8 (2.9$^{\dagger\dagger}$) \\
  \end{tabular}
\hangcaption{\footnotesize{A summary of the sensitivities when the detector bath temperature is 100~mK and the mirror/aperture temperature is 4~K. The pixel diameters are 18~mm and 12~mm for an LFP and an HFP respectively. The fractional bandwidth is 23~\% for LFPs and 30~\% for HFPs. $^\dagger$ FWHM. $^{\dagger\dagger}$Using 100 and 140~GHz bands only. }\label{tab:sensitivity}}
\end{center}
\end{table}

\section{Discussions}
\label{sec:discussion} 
The detector noise in Table~\ref{tab:sensitivity} is computed based on the assumptions of a 100~mK detector bath and 4~K mirrors and baffles. If these temperatures are higher, the detector sensitivity will degrade. While the superconducting material for the 100~mK bath temperature is manganin, we also consider NbTi and Al, with corresponding bath temperatures of 300~mK and 560~mK. We evaluate the array NET for these bath temperatures and mirror and baffle temperatures of 2, 4, 7, 10 and 30~K. Table~\ref{tab:impact} shows the array NET for these various temperature combinations. At a bath temperature of 100~mK the baffle temperature can be increased to 10~K while still meeting our requirement, while at a mirror and aperture temperature at 4~K the higher detector bath temperatures may be an option. However for higher bath temperatures the bolometer time constant slows down and systematic effects due to uncertainty in the bolometer and readout time constants has to be carefully studied~\cite{hanany,planck_timeconst}. 

\section{Conclusions}
LiteBIRD is a next-generation CMB polarization satellite that aims to measure the primordial B-mode. LiteBIRD has two detector options, TES and MKID, and here we present the focal plane design for the TES option. The multichroic TES efficiently utilizes the limited focal plan area available on a satellite platform, and we show that the LiteBIRD target sensitivity is realizable at the design level with existing technology currently being employed by ground-based experiments. LiteBIRD is currently in design phase and various studies, including evaluation of systematic effects, optical system design, thermal design, and orbit and satellite BUS system design, are ongoing. 
The first round of design work will be completed by the end of Japanese fiscal year 2013 to be ready for the mission definition review and the target launch date is 2020. 

\begin{table}[tb]
\begin{center}
  \begin{tabular}{c|c|c|c|c|c}
    &  \multicolumn{3}{c}{Aperture \& mirror temperatures}  \\ 
     Bath temperature & 2 & 4  & 7 & 10 & 30 \\ 
     $[$mK] & [K] &[K] & [K]  & [K] & [K]     \\ 
    \hline     \hline
    100 & 0.80 (1.30) & 0.97 (1.61) & 1.29 (2.07) & 1.58 (2.47) & 2.96 (4.43) \\
    \hline
    300 & 0.88 (1.43) & 1.05 (1.77) & 1.40 (2.27) & 1.72 (2.71) & 3.21 (4.84) \\
    \hline
    560 & 0.93 (1.45) & 1.15 (1.96) & 1.54 (2.52) & 1.89 (3.00) & 3.50 (5.33) \\
  \end{tabular}
\hangcaption{\footnotesize{The weighted array NET $NET_{arr}$ using all the bands for bath temperatures of 100, 300 and 560~mK and aperture and mirror temperatures of 2, 4, 7, 10 and 30~K. The numbers in parentheses are the weighted array NETs using the 100 and 140~GHz bands only. }\label{tab:impact}}
\end{center}
\end{table}

%\begin*{acknowledgements}
%This work is supported by the Ministry of Education, Culture, Sports, Science and Technology (MEXT) of Japan under KAKENHI Grant Number 21111002, 21111003. 24111715, 24740182.
%\end*{acknowledgements}

\end{document}